\documentclass[review,12pt]{elsarticle}

\usepackage{graphicx}
\usepackage{amssymb}
\usepackage{float}
\usepackage[none]{hyphenat}
\usepackage[section]{placeins}

\biboptions{authoryear}

\journal{Geothermics}

\begin{document}

\begin{frontmatter}

\title{Transverse Hydraulic Fracture Initiation: Insights from 
Intermediate-Scale Hydraulic Fracture Tests}

\author{Jeffrey Burghardt\corref{cor1} \fnref{pnnl}}
\author{Hunter Knox \fnref{pnnl}}
\author{Chris Strickland \fnref{pnnl}}
\author{Vincent Vermeul \fnref{pnnl}}
\author{Pengcheng Fu \fnref{llnl}}
\author{Craig Ulrich \fnref{lbnl}}
\author{Mark McClure \fnref{resfrac}}
\author{Mathew Ingraham \fnref{snl}}
\author{Paul Schwering \fnref{snl}}
\author{Timothy Kneafsey \fnref{lbnl}}
\author{Doug Blankenship \fnref{snl}}
\author{EGS Collab Team \fnref{collab}}

\cortext[cor1]{Corresponding author email: jeffrey.burghardt@pnnl.gov}
\address[pnnl]{Pacific Northwest National Laboratory, Richland, Washington, USA}
\address[llnl]{Lawrence Livermore National Laboratory, Livermore, California, USA}
\address[lbnl]{Lawrence Berkeley National Laboratory, Berkeley, California, USA}
\address[snl]{Sandia National Laboratories, Albuquerque, New Mexico, USA}
\address[resfrac]{ResFrac, Palo Alto, California, USA}
\fntext[collab]{Names of the entire EGS Collab team appear in the acknowledgements at the end of the paper}

\begin{abstract}
In a uniform stress field, a tensile hydraulic fracture will favor propagation in the plane formed by the maximum (most compressive) and intermediate principal stresses. For the propagation of hydraulic fractures in non-uniform stress fields, especially those where the principal stress directions change spatially, things become much more complex, are less well understood, and therefore are more difficult to predict. Such is often the case when a hydraulic fracture is initiated from a borehole drilled in the direction of the minimum principal stress. This is a well-known problem that has been the subject of several numerical and experimental studies. The purpose of this paper is to offer new insights from a series of hydraulic fracture tests conducted in sub-horizontal boreholes drilled in the Sanford Underground Research Facility. This project, called the Enhanced Geothermal Systems (EGS) Collab project, has developed a test bed at a depth of approximately 1478 meters underground consisting of sub-horizontal injection, production, and monitoring wells. This paper presents stimulation test data, and supporting geologic/geophysical characterization data. The test data are compared to previous numerical and experimental studies to determine how well this new experimental data compares with previous modeling, laboratory, and field tests.
\end{abstract}


\end{frontmatter}

\section{Introduction}
\label{section:intro}
Hydraulic fractures are an essential component of most concepts for developing enhanced geothermal systems (EGS). Early concepts of EGS were based upon highly deviated wells drilled in the direction of the minimum horizontal stress so as to produce transverse fractures, which is very similar to the modern approach used in unconventional petroleum reservoirs. This geometry has the potential to create an enormous surface area for heat transfer and concomitant large energy production rates and reservoir lifetimes \citep{Li2016}. Eventually this approach became overshadowed by the shear stimulation approach that favored vertical or slightly deviated wells drilled in the direction of the maximum horizontal stress \citep{Jung2013}. More recently, there as been renewed interest in the EGS concept centered around highly deviated/horizontals and transverse fractures \citep{Li2016, Kumar2019}.

Generating transverse fractures, however, can lead to near-wellbore fracture geometries that create large frictional pressure losses. These frictional pressure losses may have even more profound implications for EGS than they do for petroleum reservoirs because proppant degradation is accelerated in harsh EGS conditions and EGS require longer production times (decades instead of years) to achieve return on investment EGS environments than do unconventional petroleum reservoirs \citep{Li2016}. 

The reason for increased near-wellbore tortuosity in transverse fractures compared to longitudinal fractures is that the near-wellbore and far-field preferred fracture orientations are no aligned. In a uniform stress field tensile hydraulic fractures will favor propagation in the plane formed by the maximum (most compressive) and intermediate principal stresses. This is one of the fundamental guiding principles of hydraulic fracture design. For the propagation of hydraulic fractures in non-uniform stress fields, especially those where the principal stress directions change spatially, things become much more complex, are less well understood, and therefore are more difficult to predict. Such is often the case when a hydraulic fracture is initiated from a borehole drilled in the direction of the minimum principal stress. As the borehole is pressurized in this scenario (in the absence of any perforation or other defect in the borehole wall), the hoop stress is reduced while the stress in the direction parallel to the borehole is only slightly changed due to the Poisson’s effect. Eventually at some borehole pressure, the hoop stress becomes the minimum principal stress adjacent to the borehole. Thus, in the region adjacent to the pressurized borehole, the preferred direction of fracture propagation is along the axis of the borehole (i.e., a longitudinal fracture), which is incompatible with the preferred direction far from the borehole. 

As will be discussed below, this incompatibility leads to fracture behavior in the near-borehole region that is complex and difficult to predict. Furthermore, this near-borehole complexity can create a tortuous and narrow connection to the rest of the fracture, which will re-orient to match the far-field stress directions. This misalignment of the near-wellbore and far-field fracture directions can undermine the effectiveness of the created hydraulic fracture in both petroleum and geothermal settings.

In the remainder of this introductory section we will briefly discuss some past research on this topic. Then in the sections that follow, data from a set of intermediate-scale hydraulic fracture tests will be presented and discussed in the context of these overarching questions regarding about the competition between near-wellbore and far-field preferred fracture propagation directions, what factors influence the fracture geometry, and what diagnostics can be applied to understand this behavior in practice.

\subsection{Hydraulic fracture initiation}
\label{subsection:hydraulic_fracture_initiation}
Hydraulic fracture initiation has several features that differentiate it from fracture propagation in the far-field (here used to denote regions removed from stress concentrations associate with the borehole). It is prudent to first clarify a few key terms that will be used throughout this paper, as their usage has some variability even among experts in this domain.

The first pair of key terms is “fracture initiation” pressure and its difference from “breakdown” pressure. It is not unusual for these terms to be used interchangeably, but here and in other works from the literature an important distinction is made between the two \citep{Detournay1997,Zhao1996,Lecampion2015,Bunger2010,Stanchits2015}. Briefly stated, fracture initiation pressure is the pressure at which a fracture first initiates or begins to propagate. Breakdown pressure is the local maxima of the pressure vs. time curve immediately following fracture initiation. As the cited literature explains, in some situations fracture initiation has little or no expression in the pressure vs. time curve. Fast pressurization, high borehole storage (defined in the next paragraph), and high viscosity all tend to cause both an increase in breakdown pressure and a masking of any signature associated with fracture initiation in the pressure record.

Borehole storage is simply the elastic storage of fluid as the borehole is pressurized. This is a combination of compression of the fluid, expansion of the borehole, hydraulic lines, and pumping system components, and is here denoted $U_{sys}$. The magnitude of this compressibility plays an important role in the fracture initiation and breakdown process \citep{Lecampion2017,Lecampion2015}.

\subsection{Fracture initiation in deviated wells}
\label{subsection:fracture_initiation_deviated_wells}
As discussed briefly above, wells drilled in the direction of the minimum principal stress are common because of the potential they offer for large reservoir contact when multiple transverse fractures are formed. Because of the importance of this well orientation, this issue of near-borehole fracture formation has received significant attention, beginning in the 1970’s, with interest building as deviated and horizontal wells became more common.

Several studies have shown that when a borehole is drilled at an oblique angle to the principal stress directions, multiple mixed-mode fractures form at the borehole and gradually coalesce into a single dominant fracture that is oriented with the far-field stresses 2-5 borehole diameters away from the well \citep{Daneshy1973,Weijer1994,Abbas1996}. Several of these laboratory studies have shown that when a well is aligned with the minimum principal stress both longitudinal and transverse fractures form at the borehole. The longitudinal fractures extend only a few borehole diameters away while the transverse fractures extend into the far-field. Although image logs are seldom used in highly deviated or horizontal wells, a few field studies have observed this combination of longitudinal and transverse fractures \citep{Waters2006,Jeffrey2015}. The \citet{Weijer1994} study found that higher pressurization rates and higher fluid viscosity tended to promote the formation of both transverse and longitudinal fractures, while lower pressurization rates and viscosity tended to favor the formation of only transverse fractures. Note that most of the laboratory studies mentioned thus far used nominally cylindrical/conical perforations that are perpedicular to the borehole axis, as are created with shape charge perforating guns.

These and other laboratory and field studies have shown that breakdown pressure generally increases monotonically as the direction of a well changes from alignment with the intermediate to the minimum principal stress \citep{Abbas1996,Veeken1989,Owens1992}. A summary of these reorientation studies can be found in \citet{Abbas2013}.

\section{Test site description}
\label{section:test_site_description}

The Enhanced Geothermal Systems (EGS) Collab project provides a new data set from a series of hydraulic fracture tests conducted in sub-horizontal boreholes drilled in the Sanford Underground Research Facility (SURF) in the Black Hills region of South Dakota. EGS Collab has developed a very well characterized test bed at a depth of approximately 1478 meters underground by drilling sub-horizontal boreholes from a mine drift. The stimulated borehole was drilled in the direction of the anticipated minimum principal stress and radial 'notches' that were approximately 6 mm deep were created at the planned stimulation locations.

This test design allowed relatively large hydraulic fractures to be created in highly-stressed rock with relatively short boreholes that allowed for a high degree of characterization via logs, core, and geophysical monitoring from an array of nearby monitoring boreholes.

This paper will present test data and supporting geologic/geophysical characterization data. The test data are compared to previous numerical and experimental studies to determine how well this new experimental data compares with previous modeling, laboratory, and field tests. The data and results presented here are intended to contribute to the body of literature on transverse fracture initiation.

A brief overview of the test site will be given here. For further details the reader is referred to \citet{Kneafsey2019}, and the citations therein. All tests described in this paper were performed within the Poorman phyllite formation. Based on these previous tests the stimulation well for the tests described below were drilled with an azimuth of approximately $0^{\circ}$ with an inclination of $78^{\circ}$. Well surveys and plots of all well trajectories can be found at \citet{Neupane2018}.

The Poorman is well foliated, generally having distinct bands that are 1-5 cm thick, which are thought to be remnants of bedding from the protolith \citep{Caddey1991}. This location is on the West limb of the southeast-plunging Lead anticline. Besides the large-scale Lead anticline, there are numerous folds ranging in scale from tens of meters to less than a meter. Initial electrical resistivity tomography results suggest that there is a fold with a length scale of ~10 m running through the test bed \citep{Johnson2019}. The nominal temperature of the testbed is $30-35^{\circ}$ C \citep{White2019}.

These numerous folds, along with multiple natural fracture sets and other heterogeneities, make the structure of this test bed quite complex. Despite the presence of these features, over the tested interval ranging from 39 m (128 ft) to 50 m (164 ft), the strike and dip of the foliations were fairly constant, with a few small-scale folds causing major deviations from the broader trend. Picks from image logs show an overall trend with a strike of S$40^{\circ}$W and a dip of approximately $84^{\circ}$.

A considerable amount of geomechanical and hydrological testing has been performed at the SURF facility from the 1970’s until the present. \citet{Pariseau1986} compiled stress measurements made up to 1986 and formulated a mine-scale regression for the state of stress, in MPa, as a function of vertical depth, $z$, in kilometers is:

\begin{equation}
    \label{eq:sigz}
    \sigma_V = 28.3 z
\end{equation}
\begin{equation}
    \label{eq:shmax}
    \sigma_H = 14.3 + 12.0 z
\end{equation}
\begin{equation}
    \label{eq:shmin}
    \sigma_h = 0.8+12.4 z
\end{equation}
where $\sigma_V$, $\sigma_H$, and $\sigma_h$ are the vertical, maximum horizontal, and minimum horizontal principal stresses, respectively.

Most recently, hydraulic fracture-based stress measurements were made adjacent to the EGS Collab site for the kISMET project \citep{Oldenburg2016, Wang2017}. This location is approximately 20 meters away from the EGS Collab site, with wells drilled vertically from the invert (floor). Stress measurements in these boreholes found a normal-faulting stress state with a minimum principal stress between 20.0 and 24.1 MPa at a total vertical depth (TVD) between 1520 and 1550 m \citep{Wang2017}. The fractures induced in these tests indicate a minimum compressive stress with a strike of N$2^{\circ}$E (corrected for magnetic north) with a downward plunge of $9.3^{\circ}$ \citep{Kneafsey2020, Schwering2020}.

\section{Test procedures}
\label{section:test_procedures}

The results of three stimulation tests will be described, and will be referred to by the date on which they were conducted: May 21, May 22, and July 18. As mentioned in Section \ref{section:intro}, fluid pressure does not generate significant change in stress acting parallel to the borehole, longitudinal fractures are likely to form in the absence of any significant defects in the borehole wall. Because of this it was decided to create circumferential notches in the borehole at the target stimulation locations to help initiate a transverse hydraulic fracture. The measured depths and start dates for the three stimulation tests are given in Table \ref{tab:stimulations}. 

The notch at each location was created with a specialized tool developed for this project. This tool uses a cutting bit mounted to the end of the drill string. The cutter was designed to create a notch with a depth of 6 mm. For each test a straddle packer assembly was positioned with the notch located within the packed-off interval (hereafter referred to simply as “the interval”). The interval length for these tests was 1.65 m (5.4 ft).

The straddle packer assembly was outfitted with two hydraulic lines connected to the interval. One of these lines was used to pump water to generate the fracture. For the May 21 stimulation pressure was only measured at the pump outlet, which will inevitably be somewhat higher than the actual interval pressure because of frictional losses. However, for the low injection rates used in the operations discussed here the frictional losses are less than 10 psi. For May 22 and July 18 stimulations the other line was used as a non-flowing pressure measurement line. This allowed more accurate measurement of the interval pressure without the effects of frictional pressure losses. Pressure on the non-flowing line was measured using a transducer with a full-scale range of 68.6 MPa (10,000 psi) and accuracy of 0.2\% of full-scale range, or $\pm 0.14$ MPa (20 psi).

A Chandler Engineering Quizix Q6000-5K pump was used for all injections reported here. This is a positive-displacement dual-cylinder syringe pump. The injection rates reported here were all measured with the control system of this pump, which has a nominal accuracy of 0.001 mL/min. The temperature of the injected water during the stimulations was approximately $29^{\circ}$ C.

During two of the three tests reported here a borehole displacement measurement tool called Step-Rate Injection Method for Fractures In-situ Properties (SIMFIP) was used \citep{Kakurina2020}, though the details of those measurement will be described in a separate publication. This tool requires a certain level of fluid pressure in the interval to set its clamps. Therefore, the borehole was slowly pressurized to 6.9 MPa (1000 psi), and maintained at this pressure while these clamps were engaged. After the clamps were engaged the fracture was initiated at a constant pumping rate.

For each test the general procedure was to pump approximately 2 L of water into the initiated fracture and then shut in the interval to allow the fracture to gradually close. The data during the shut-in period were used to estimate the fracture closure pressure, leakoff coefficient, and fracture toughness \citep{Economides2000}. 

This fluid volume was chosen because it is the volume that was estimated to drive an idealized radial fracture, based on best estimates for rock properties from core measurements and previous tests at the kISMET site, to a radius of 1.5 meters. Following this first shut-in period, the fracture was then driven with a volume of fluid estimated to create a fracture radius of 5 meters, followed by another shut-in period. These portions of the test are beyond the scope of this paper. For the May 21 and May 22 stimulations fractures were eventually driven until they intersected a sub-parallel production well. However, this paper will focus only on the fracture initiation and driving to an idealized radius of 1.5 meters, which took place on the dates shown in Table 1. More comprehensive testing on or from these notches was performed over several months in 2018 and 2019, which will be discussed in other publications.

For each test the slope of the linear portion of the pressure vs. time plot was calculated and used to compute the borehole storage compressibility. The point at which the pressure vs. time curve begins to depart downward from this line is taken as an indication of fracture initiation. The instantaneous cumulative volume $V_{frac}$ injected into the fracture is calculated based on the measured compressibility using:
\begin{equation}
    \label{eq:vfrac}
    V_{frac}=V_{pump}-U_{sys} P_{inj}
\end{equation}
where $V_{pump}$ is the pumped volume and the instantaneous flow rate is obtained by taking the derivative of Eq. \ref{eq:vfrac} with respect to time (de Pater et al., 1994):
\begin{equation}
    \label{eq:qfrac}
    Q_{frac}=Q_{pump}-U_{sys}  (dP_{inj})/dt
\end{equation}
It is easy to see from this expression that when the pressure is constant, there is no change in the borehole storage volume with time and the flow rate into the fracture exactly equals the pump rate.

\begin{table}[h]
\centering
\begin{tabular}{l c}
\hline
\textbf{Measured Depth m (ft)} & \textbf{Date of First Stimulation} \\
\hline
 43.2 (142) & 5/21/2018  \\
 50 (164) & 5/22/2018 \\
 39 (128) & 7/18/2018 \\
\hline
\end{tabular}
\caption{ \label{tab:stimulations} Measured depths and start dates for the three stimulation tests}
\end{table}

\section{Test results}
\label{sec:test_results}

\subsection{43 m notch stimulation}
\label{subsec:stimulation_1}
\begin{figure}[!htb]
\centering\includegraphics{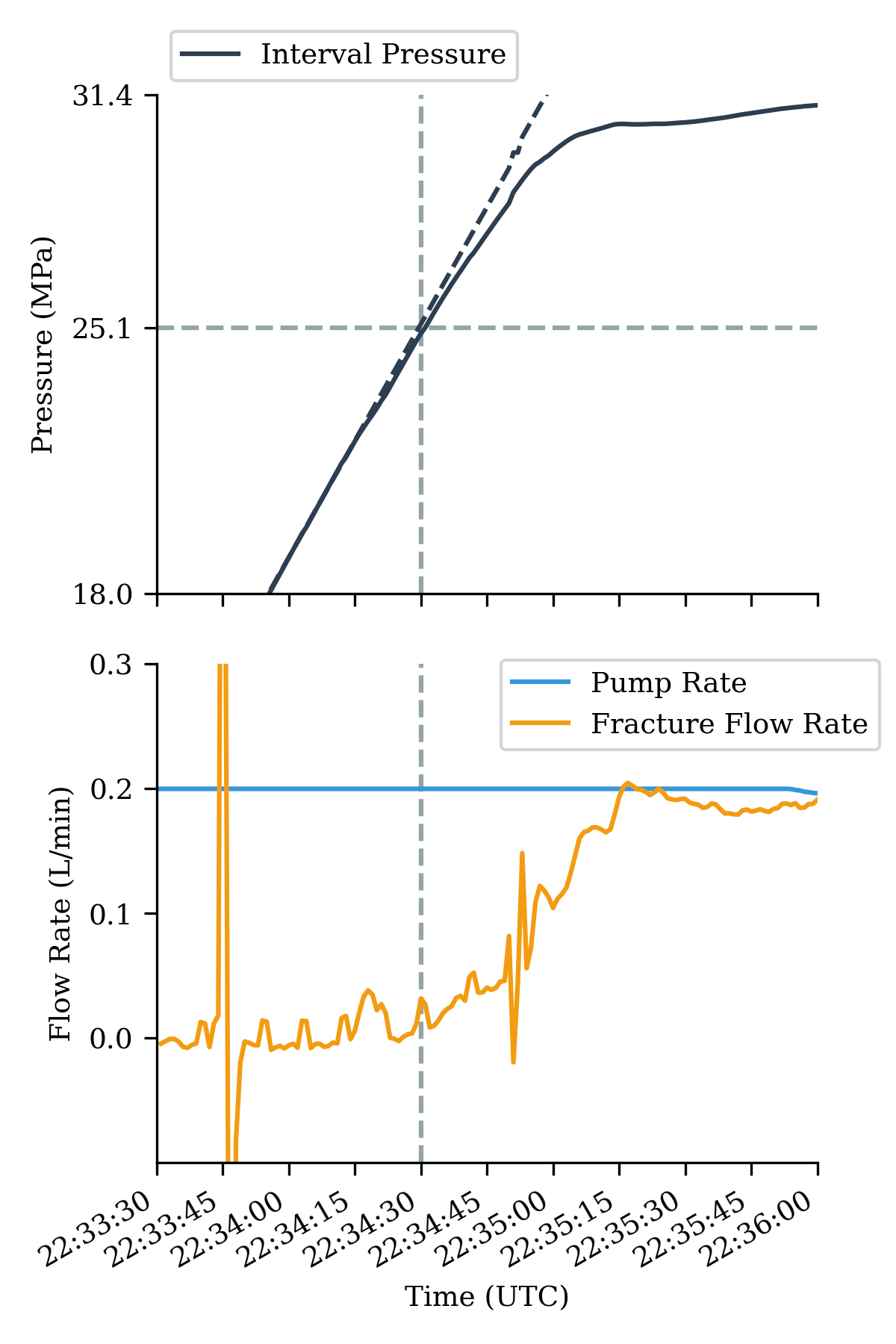}
\caption{\label{fig:fig1}Plot of pressure and volume data during breakdown of the May 21, 2018 stimulation}
\end{figure}

Figure \ref{fig:fig1} shows a plot of the data acquired during the breakdown phase of the May 21 stimulation of the notch located at 43.2 m (142 ft). The slope of the linear portion of the injection pressure vs. time curve, together with the pump rate of 200 mL/min, was used to calculate a system compressibility of 62.7 MPa/L. The fracture initiation pressure, indicated by the horizontal dashed line shows where the pressure vs. time curve deviates downward from the linear trend, was estimated to be 25.1 MPa (3646 psi).
As can be seen in Figure \ref{fig:fig1}, the injection pressure continued to rise considerably following fracture initiation, with no significant local maximum and subsequent decline as is often observed in hydraulic fracture initiation. 

\begin{figure}[!htb]
\centering\includegraphics{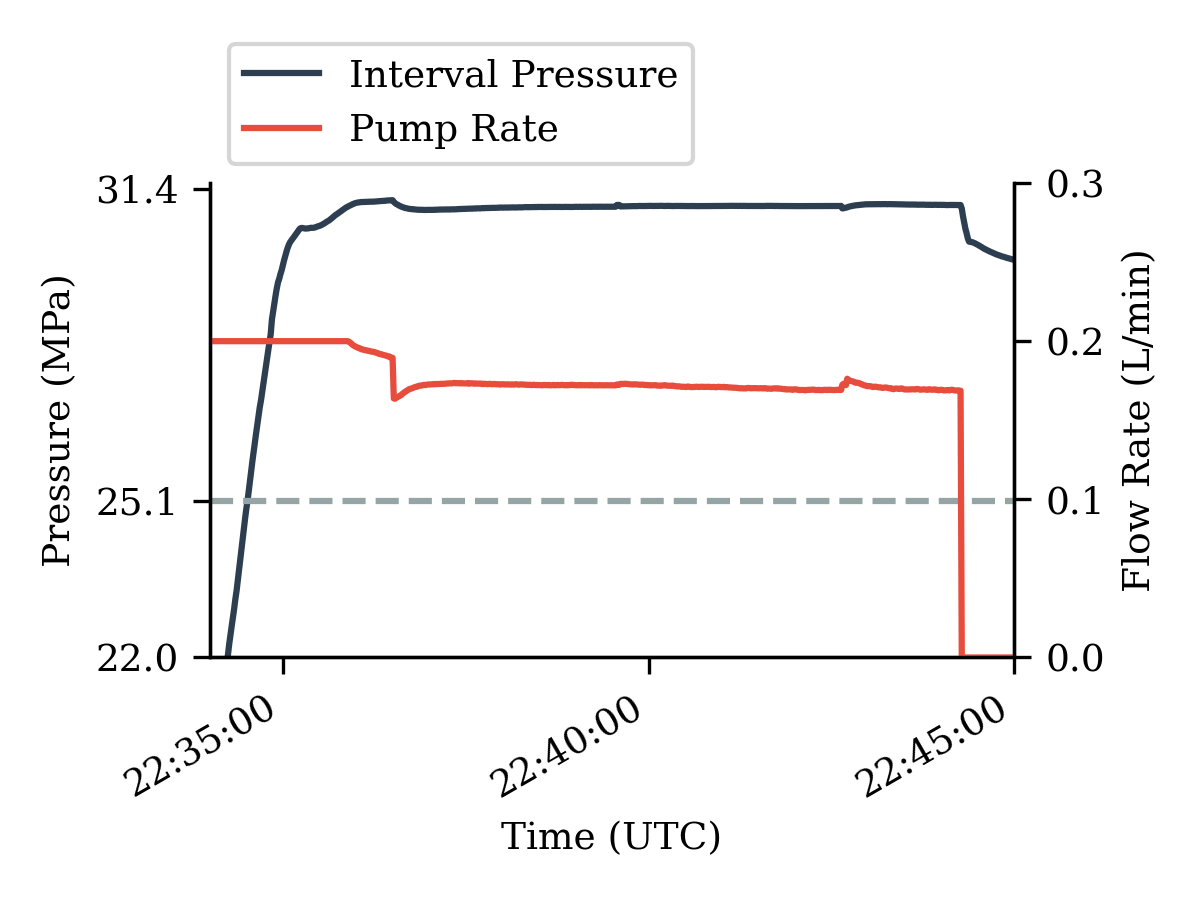}
\caption{\label{fig:fig2}Plot of data during entire 1.8 L total injection for the May 21, 2018 stimulation}
\end{figure}

Figure \ref{fig:fig2} shows a plot of the data during the entire 1.8 L injection. The injection rate was reduced approximately 1/3 of the way through the injection because a pressure limit of 31 MPa had been set in the pump’s control system. When the injection pressure approached this value the pump rate was automatically reduced so that the limit would not be exceeded.

\begin{figure}[!htb]
\centering\includegraphics{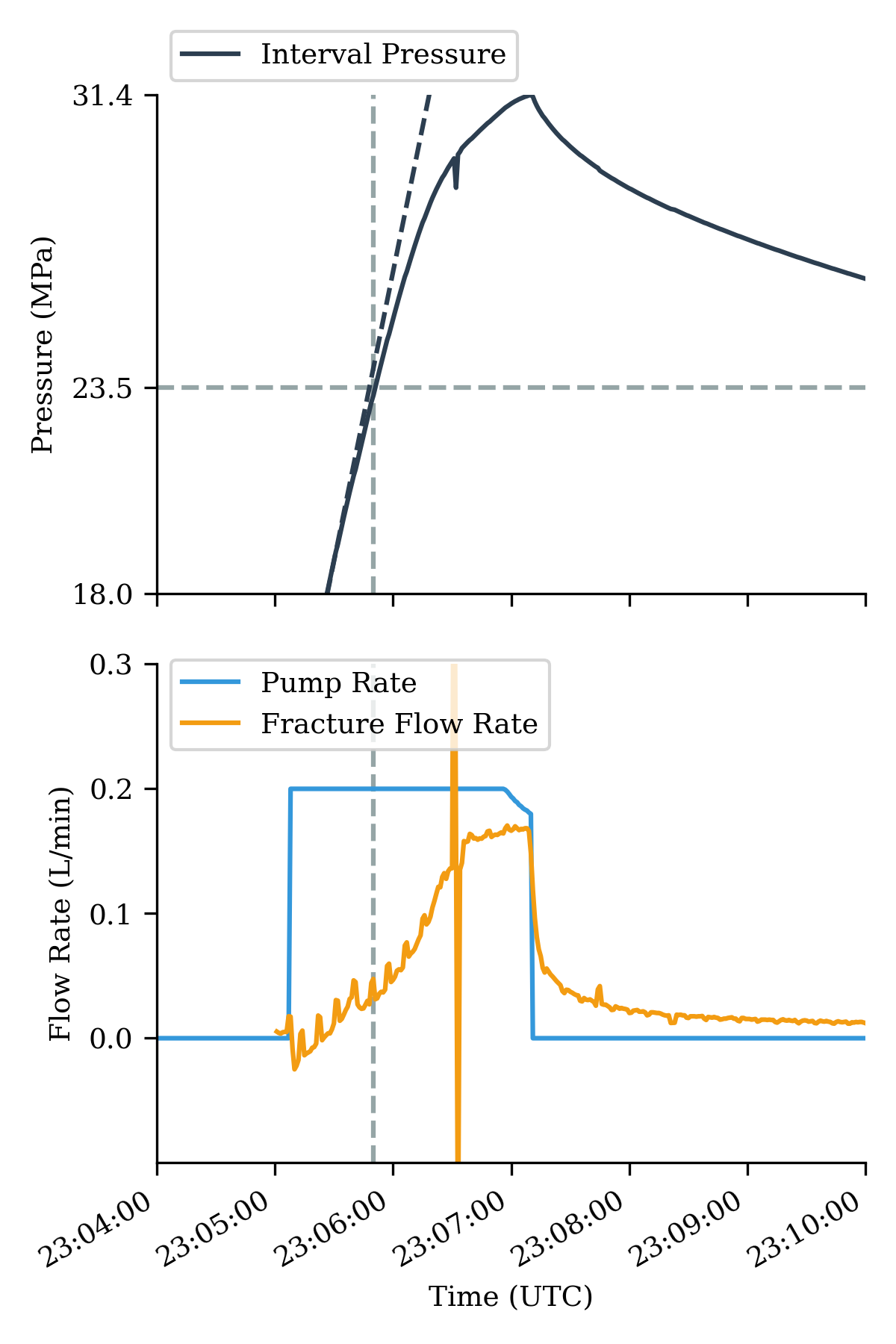}
\caption{\label{fig:fig3}Plot of data during reopening test of the May 21, 2018 stimulation}
\end{figure}

After a brief shut-in period a leak developed from a valve that caused venting of the interval pressure, making the data difficult to interpret. To generate a good long-term leakoff pressure data set, after the valve was fixed the interval was pressurized again at a constant rate until the fracture was evidently reopened, and the interval was again shut in. This provided an opportunity to observe the reopening response of the fracture. Figure \ref{fig:fig3} shows a plot of the data acquired during this reopening test. The reopening pressure, again interpreted from the deviation from linearity of the pressure vs. time curve, was 23.5 MPa (3412 psi), which is 1.6 MPa lower than the estimated fracture initiation pressure. Following the fracture reopening the pressure again climbed to the 31 MPa (4500 psi) pressure limit that caused the pump to reduce the injection rate.

After the fracture was reopened the well was shut in overnight. The pressure declined to atmospheric pressure after less than one hour, which is much faster than would be expected based on the very low permeability of the rock and no apparent leaks in the hydraulic system. A later stimulation was conducted at this location on December 20 and 21, 2018. During that test leakage past the upper packer was observed. The packer was moved up and down the well until a location was found where only a small leak past the upper packer occurred. 
 
\begin{figure}[!htb]
\centering\includegraphics[width=0.8\textwidth]{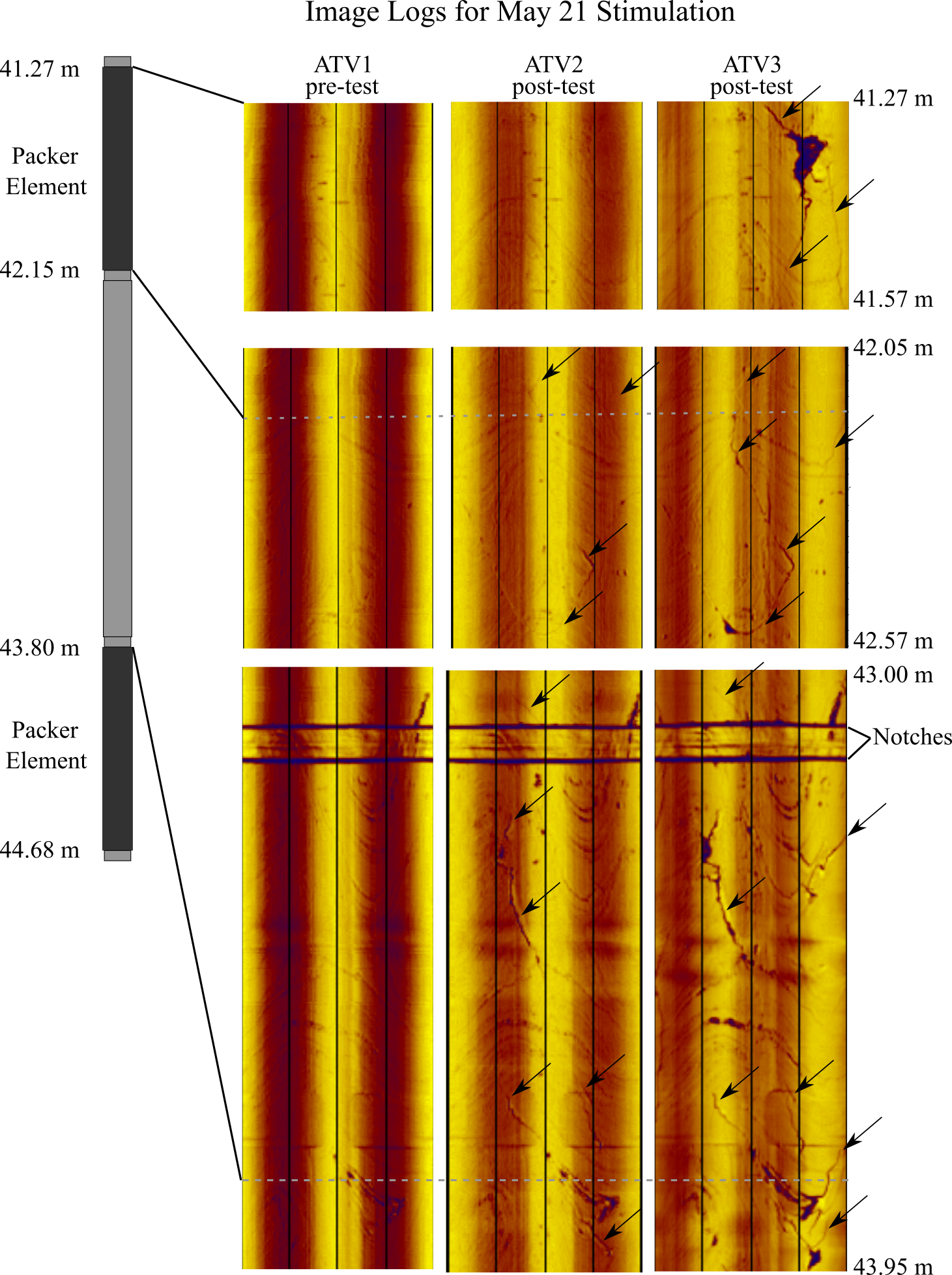}
\caption{\label{fig:fig4}Acoustic (ATV) image logs collected over the zone stimulated on May 21; ATV1 was collected prior to the stimulation, ATV2 was collected shortly after the stimulation, and ATV3 was collected in January 2020 after subsequent higher pressure stimulations at this zone; the left most edge of each log is magnetic North, which is within a few degrees of the top of the borehole }
\end{figure}

Figure \ref{fig:fig4} shows acoustic televiewer (ATV) logs that were collected in the borehole both before the stimulation (ATV1), in late May 2018 (ATV2), and again in January 2020 (ATV3). Induced fractures are identified with the thin black arrows and the two notches present in the zone are also identified. A schematic of the straddle packer system along with key alignment depths are shown in this figure for reference. These induced fracture are all sub-longitudinal to the borehole and are located primarily on the top and bottom of the sub-horizontal borehole within the interval. The induced fractures do not appear to be aligned with either the foliations or any obvious natural fracture. There are also indications of a sub-longitudinal fractures extending underneath the packers, most prominently in ATV3. These fractures were likely the cause of the rapid decline in the overnight shut-in test, and subsequent observations of leakage past the upper packer. Since the pressure rapidly declined well below any of the in situ stresses it appears that the fracture retained considerable conductivity even when mechanically closed.

The fracture system initiated in the May 21 stimulation of the 43.2 m notch was eventually driven until it intersected a production well and multiple monitoring wells a short distance away. These intersections appear to actually be multiple closely-spaced (1-2 m) fractures rather than a single fracture \citep{Schwering2019}. The intersection points with the monitoring and production wells and the cloud of localized microseismic events \citep{Schoenball2019} suggest that the strike of the primary hydraulic fracture plane is N$89^{\circ}$E and the dip is $65-80^{\circ}$. 

The intersection points at these wells suggests that that the azimuth of the far-field stress was both relatively uniform between these wells and consistent with observations from the kISMET project. However, observations in both the monitoring and production wells indicate multiple closely-spaced fractures at the intersection point with these wells.

\subsection{50 m notch stimulation}
\label{subsec:stimulation_2}

\begin{figure}[!htb]
\centering\includegraphics[width=0.5\textwidth]{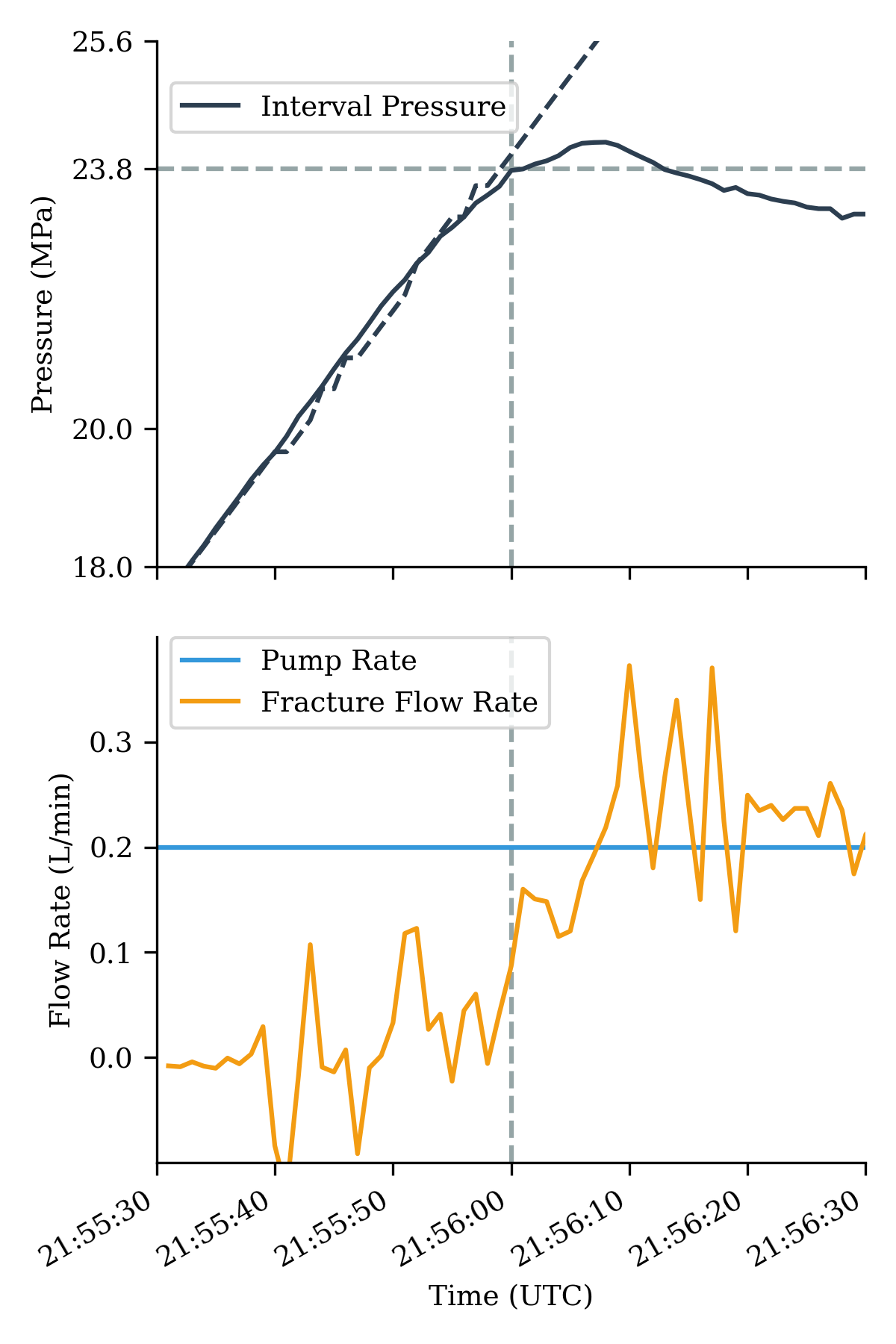}
\caption{\label{fig:fig5}Plot of pressure and volume data during breakdown of the May 22, 2018 stimulation}
\end{figure}

Figure \ref{fig:fig5} is a plot of the data acquired during the fracture initiation phase of the May 22, 2018 stimulation at the notch located at 50 m (164 ft).  The system compressibility during the pressure building phase of this test was estimated to be 70.5 MPa/L, and the fracture initiation pressure is estimated to be 22.5 MPa (3272 psi). Following fracture initiation the injection pressure rose to a peak of 24.2 MPa (3509 psi), and then declined to 23 MPa, then steadily rose to 25.4 MPa at the end of the injection. Evidenced in the lower plot of this figure, as the borehole pressure declined following the peak pressure, the flow rate of fluid into the borehole slightly exceeds the pumping rate momentarily due to borehole storage effects.
 
A total of approximately 2.1 L of water was injected for the May 22 stimulation before the interval was shut in. An analysis of the pressure transient during the shut in period was made to estimate the fracture closure pressure, radius, toughness, and leakoff coefficient \citep{Nolte1988,Economides2000}. This interpretation resulted in an instantaneous shut-in pressure (ISIP) of 24.5 MPa (3560 psi) and a closure pressure of 21.8 MPa (3160 psi), for a net pressure of 2.7 MPa (392 psi). The analysis also estimated a radius of 2 m (6.6 ft), and a leakoff coefficient of $5.5\times10^{-7}$ $\frac{\textrm{m}}{\sqrt{\textrm{s}}}$. The leakoff coefficient calculation assumes plane strain modulus of 75 GPa. The implied fracture toughness is 4.4 $\mathrm{MPa}-\sqrt{\mathrm{m}}$.

\begin{figure}[!htb]
\centering\includegraphics[width=0.5\textwidth]{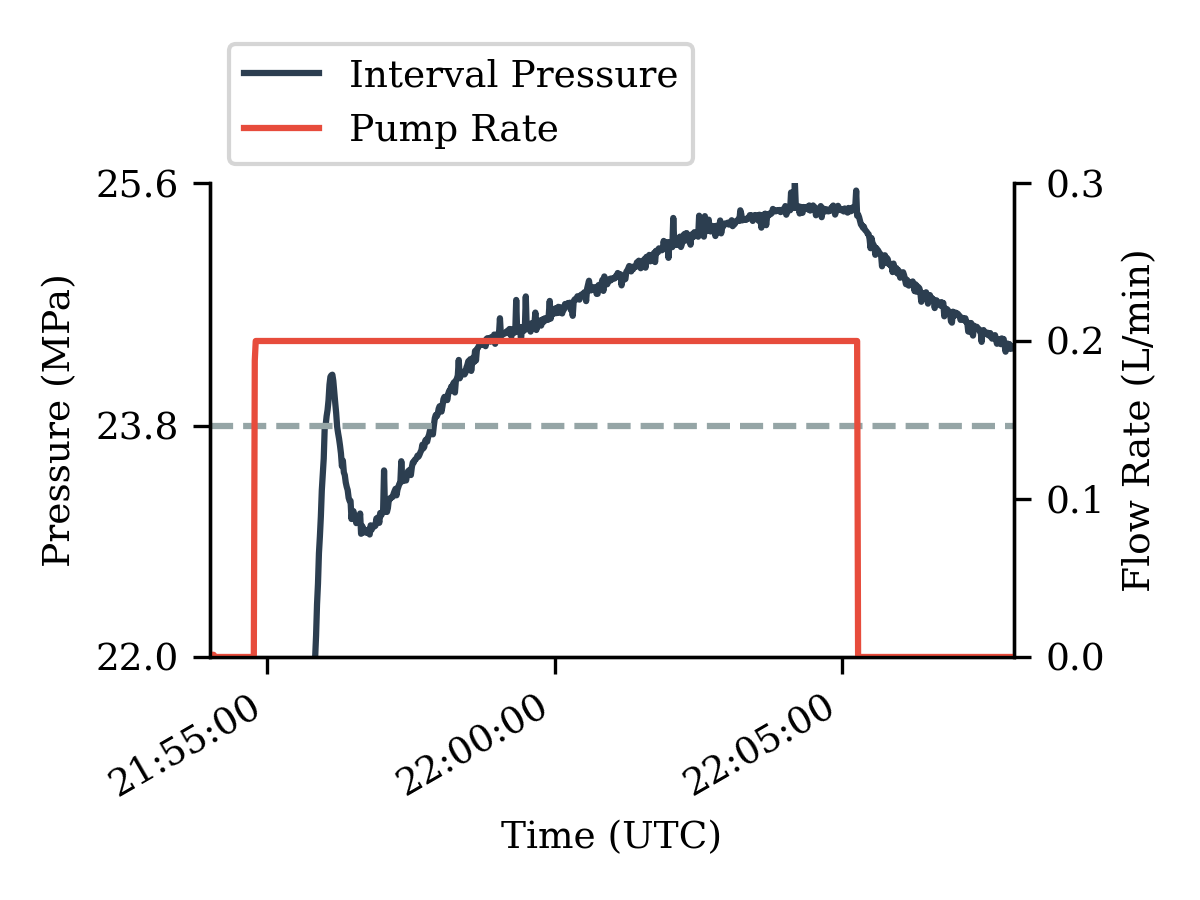}
\caption{\label{fig:fig6}Plot of data during entire 1.8 L total injection for the May 22, 2018 stimulation}
\end{figure}

\begin{figure}[!htb]
\centering\includegraphics[width=0.8\textwidth]{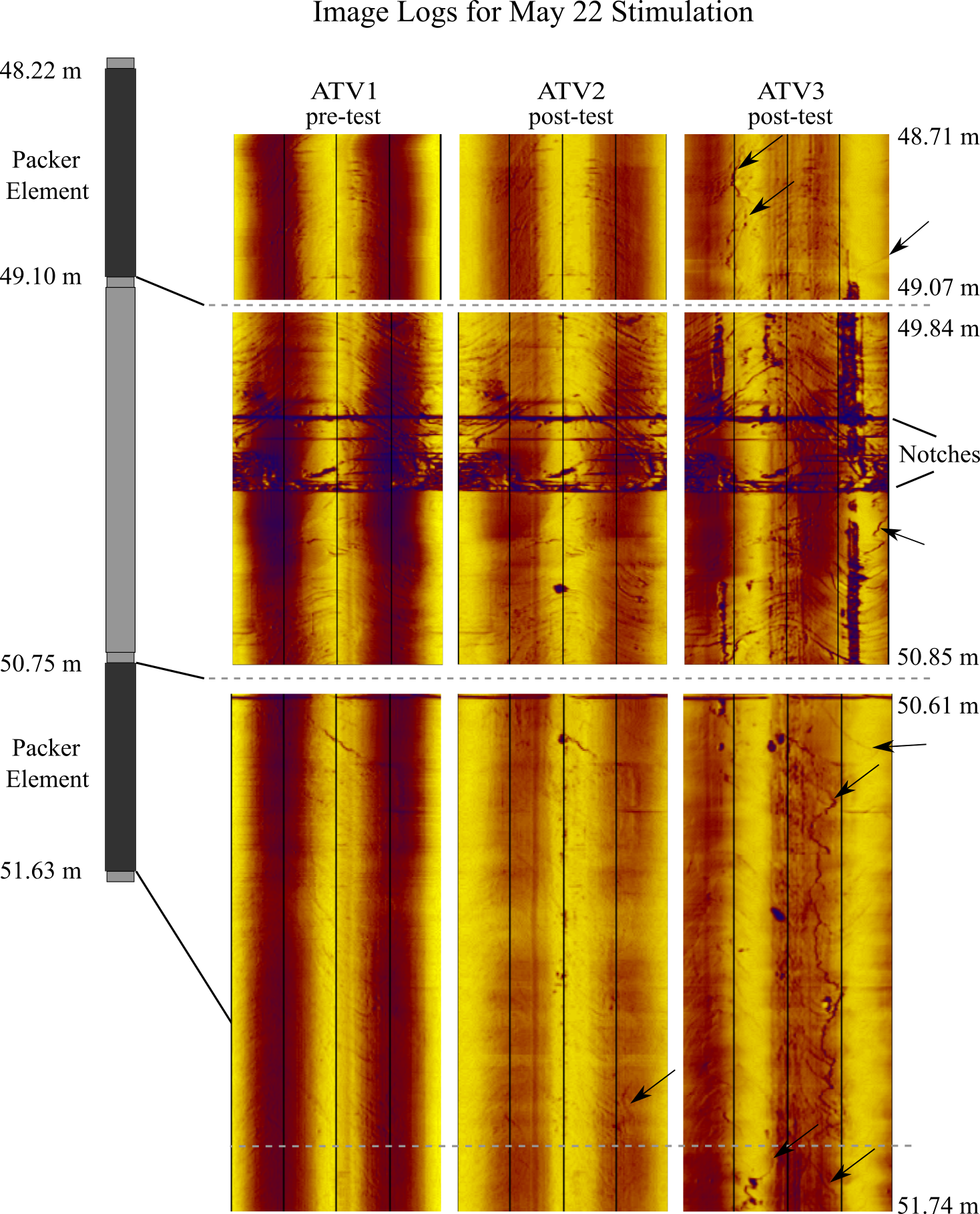}
\caption{\label{fig:fig7}Acoustic (ATV) image logs collected over the zone stimulated on May 22; ATV1 was collected prior to the stimulation, ATV2 was collected shortly after the stimulation, and ATV3 was collected in January 2020 after subsequent higher pressure stimulations at this zone; the left most edge of each log is magnetic North, which is within a few degrees of the top of the borehole}
\end{figure}

Following the initial injection of 2.1 L, a step pressure/rate test was conducted. This, and subsequent similar tests was performed by first using the pump in a constant pressure mode until a constant flow rate was achieved. Once the required flow was above approximately 50 mL/min the test continued as a step-rate test with a constant injection rate. Following this step pressure/rate test a constant rate injection at 0.4 L/min was performed until an additional 23.5 L at been injected, which was anticipated to produce a fracture radius of approximately 5 m (16.4 ft). The well was then shut in overnight. The next day another step pressure/rate test was performed followed by another period of constant rate injection at 5 L/min until the fracture had intersected the production well, after which another step pressure/rate test was performed. 

\begin{figure}[htb]
\centering\includegraphics[width=0.5\textwidth]{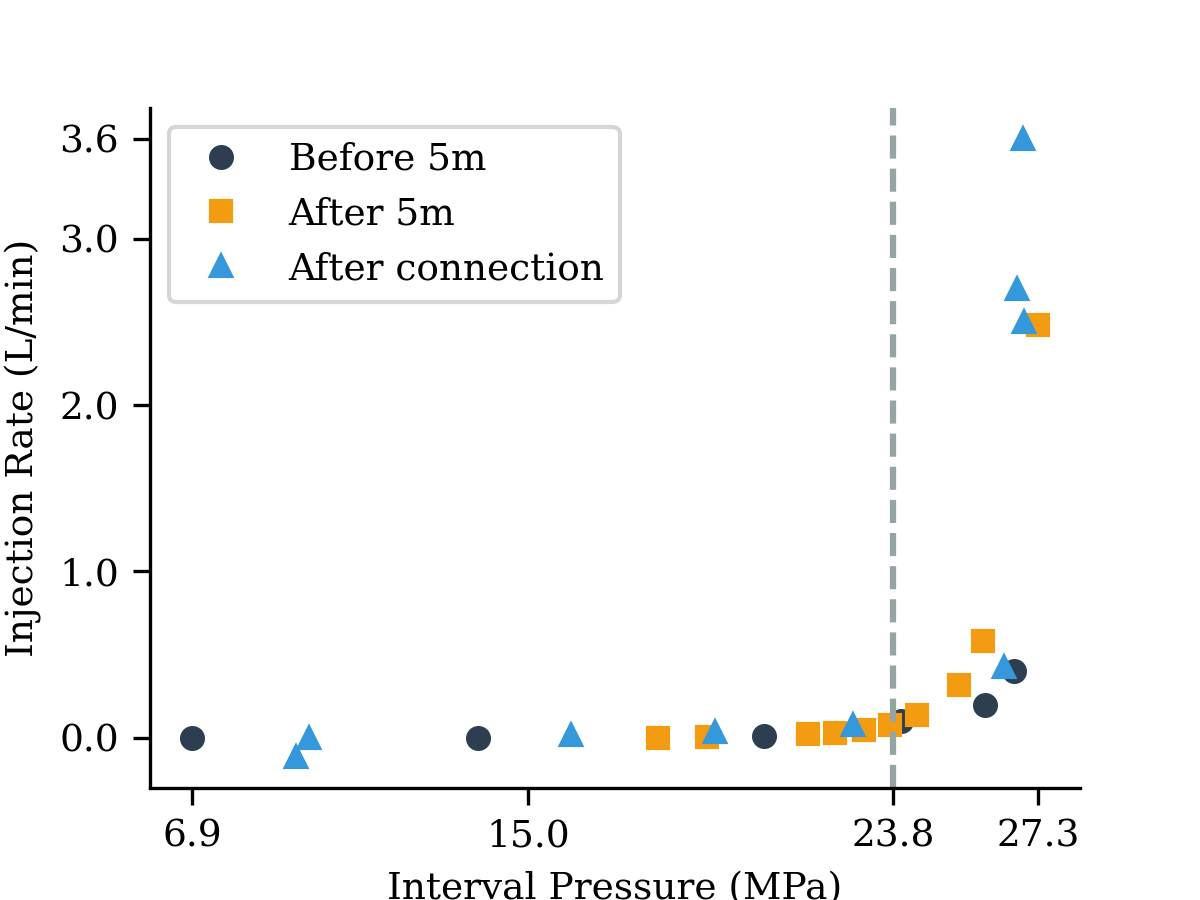}
\caption{\label{fig:fig8}Results of multiple step-rate tests on the May 22, 2018 stimulation}
\end{figure}

Figure \ref{fig:fig8} shows the results of these three step pressure/rate injection tests. In this figure ``After connection`` means after the connection to the production well was made. A larger pump was used for the “after 5m” and “after connection” step-rate tests. Owing to the larger full-scale range, the uncertainty in the flow measurements in these tests is larger (approximately ± 50 mL/min). Because of this the lower rate measurements have a larger uncertainty.

The results of these three step-rate tests are largely consistent with the analysis of the pressure transient analysis following the initial 2.1 L injection. Specifically, significant flow into the fracture occurs at approximately 22-23 MPa (3192-3336 psi), which is close to the picked closure pressure of 21.8 MPa (3162 psi). Also, a vertical asymptote occurs in the data at approximately 27 MPa (3916 psi), which is an indication of the fracture propagation pressure. This leads to an estimated net pressure of 4-5 MPa (580-725 psi), which is slightly more than the pressure transient analysis estimate. During the shut-in period there is very little flow into the fracture, so friction losses due to tortuosity do not play a role. Therefore, a reasonable interpretation is that the difference between the estimate of the net pressure from the step-rate test and the shut-in transient test is due to near-borehole tortuosity friction loss of 1.3-2.3 MPa.

\citet{Fu2021} provides a comprehensive overview of the many observations that inform the large-scale structure of the fracture induced from the 50 m notch. A brief overview of the most salient observations will be provided here. Observations in both the intersected monitoring well and production well indicate multiple closely-spaced fractures at the intersection point with these wells. \citet{Fu2021} fit five fracture planes to the intersection and microseismic data. Of these fractures the primary macro-scale fracture most closely connected to the stimulated notch at 50 m is the one labeled by \citet{Fu2021} as W-N. This fracture plane has a strike of N$72^{\circ}$E and a dip of $81^{\circ}$. Despite the apparently mostly linear/planar nature of this most closely connected fracture, the intersection with the production well reveals a great deal of small-scale complexity.

An acoustic televiewer (ATV) log was collected in late May, 2018 after the initial stimulation discussed above. Another ATV log was collected in January 2020 after a long term thermal circulation test had been conducted from this notch. Figure \ref{fig:fig7} shows a comparison of the pretest OTV and ATV logs and the two post-test ATV logs. The details of the long term thermal circulation test is beyond the scope of this article but it is mentioned since it likely had an impact on the extension of the fractures observed between the May 2018 and January 2020 ATV logs shown in Figure \ref{fig:fig7}.

Unlike the May 21 stimulation, no induced fractures are evident in the image logs collected at the end of May 2018 with a possible exception of one under the lower packer. After the May 2018 ATV log was collected it was concluded that, as with the \citet{Jeffrey2015} study, the lack of an apparent induced fracture meant that the fracture initiated from the notch, which would not be expected to create any signature on the image log. Alternatively, it could not be ruled out that the fracture opened/sheared a plane of weakness in the rock and closed in a way that left no signature on the image log. 

The image log collected in January 2020 after the long term thermal test at this location offers additional insight. There is what appears to be precipitated minerals on this portion of the well from the long term test that obscure portions of the January 2020 image log. This is believed to be what caused the long thick streaks just above and below the notches seen in ATV3 in Figure \ref{fig:fig7}. Beyond this precipitate, one of the most apparent differences observed in the January 2020 image log is the obvious longitudinal fractures underneath both the upper and lower packers. Only a small section of the fracture below the lower packer can be seen in the May 2018 image log, so the formation of these fractures likely occurred during later higher pressure testing. En echelon fractures can be observed on the January 2020 image log at 50.3 m. These seem to be aligned with the foliations, but some ligaments of the fracture extend across foliations. Some opening of foliations also appears likely at 50.5 m. This, combined with the concentration of the precipitated minerals at this location, suggests that the primary outflow of water during the long term thermal test was through the foliation planes above and below the notches.

\subsection{39 m notch stimulation}
\label{subsec:stimulation_3}

\begin{figure}[htb]
\centering\includegraphics[width=0.5\textwidth]{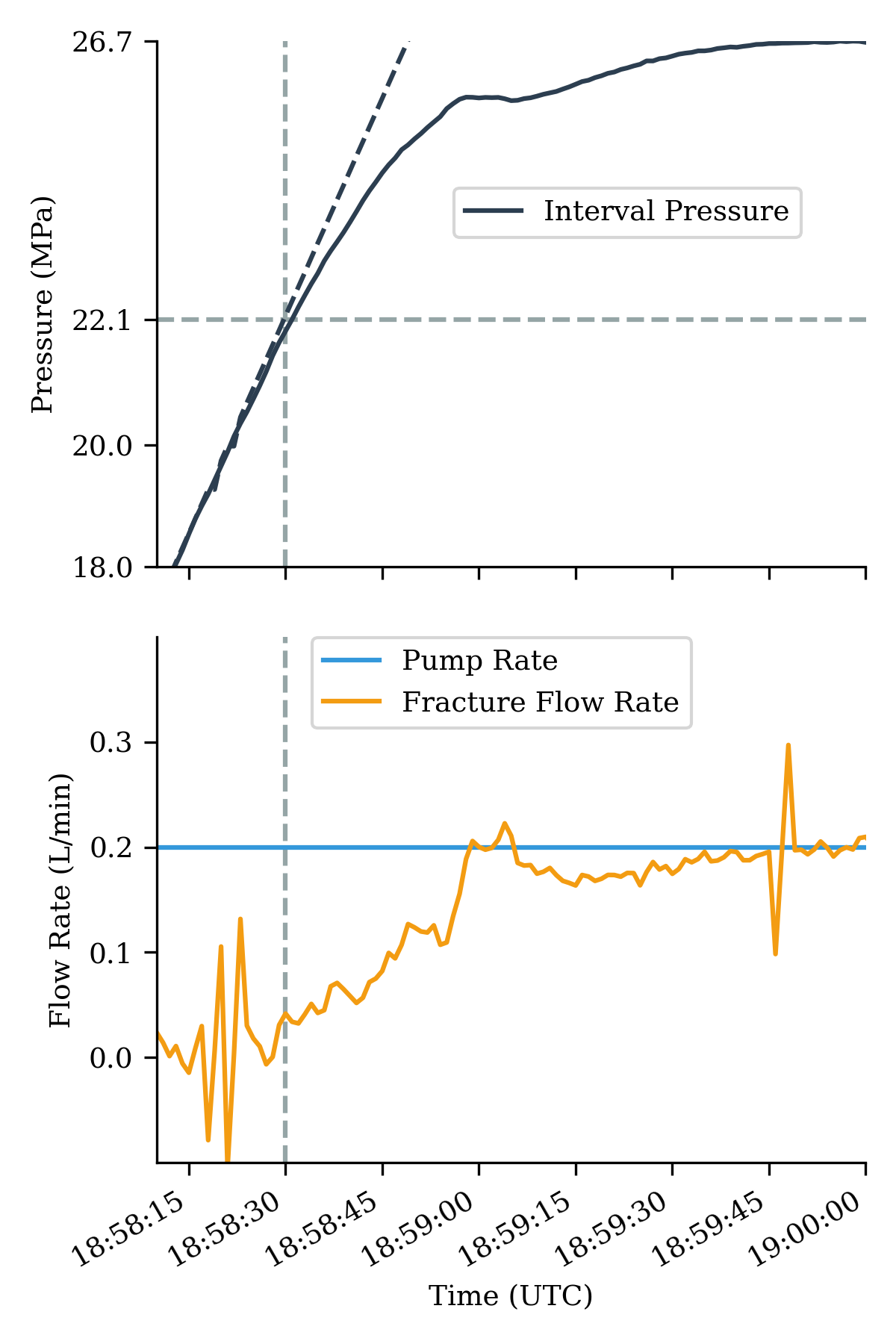}
\caption{\label{fig:fig9}Plot of pressure and volume data during breakdown of the July 18, 2018 stimulation}
\end{figure}

\begin{figure}[htb]
\centering\includegraphics[width=0.5\textwidth]{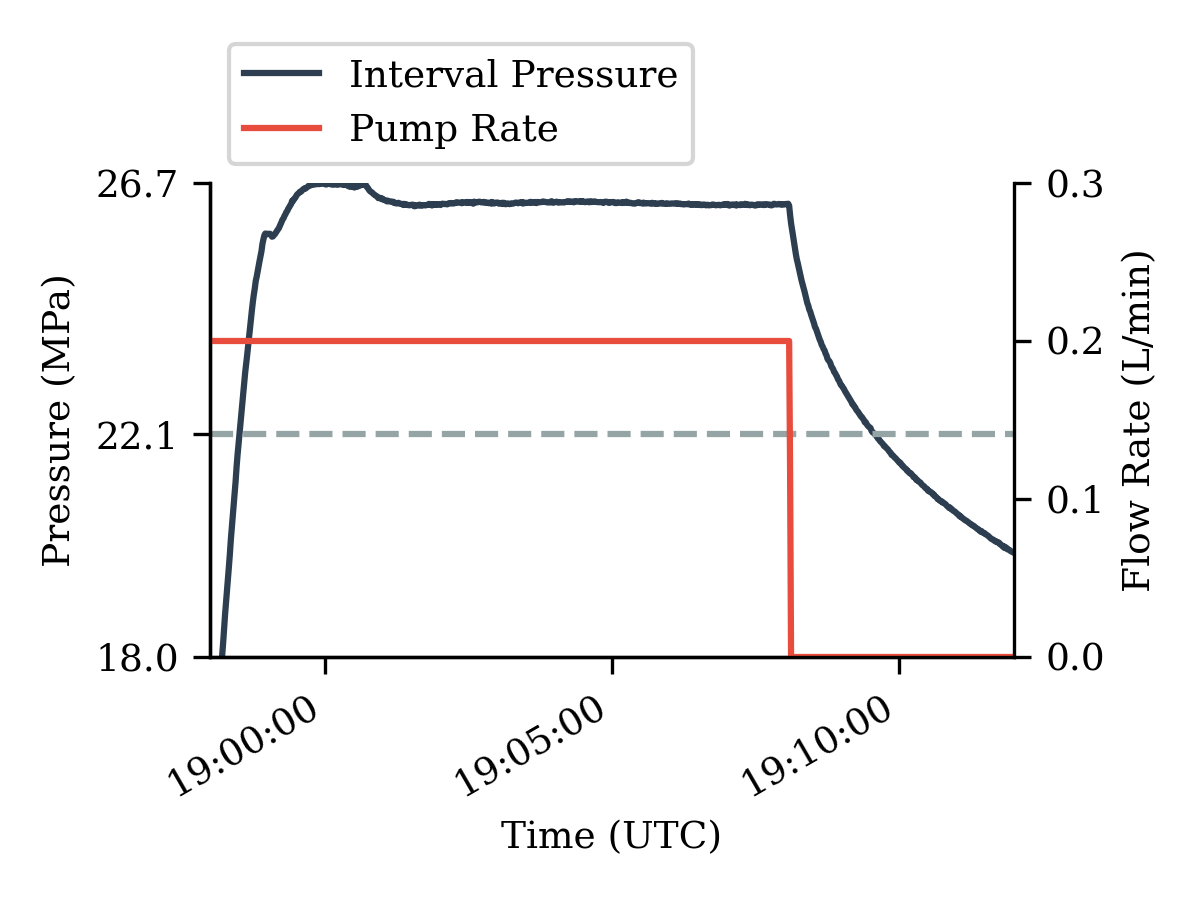}
\caption{\label{fig:fig10}Plot of data during entire 1.8 L total injection for the July 18, 2018 stimulation}
\end{figure}

Figure \ref{fig:fig9} is a plot of the pressure and flow data acquired during the fracture initiation phase of the July 18, 2018 stimulation at the notch located at 39 m (128 ft). Figure \ref{fig:fig10} shows the data during the entire ~2 L injection.  Fracture initiation occurred at 22.1 MPa (3201 psi), after which the pressure continued to increase, reaching a momentary peak at 25.7 MPa (3272 psi), followed by a brief decline, and another rise to approximately 26.6 MPa (3858 psi).

\begin{figure}[htb]
\centering\includegraphics[width=0.8\textwidth]{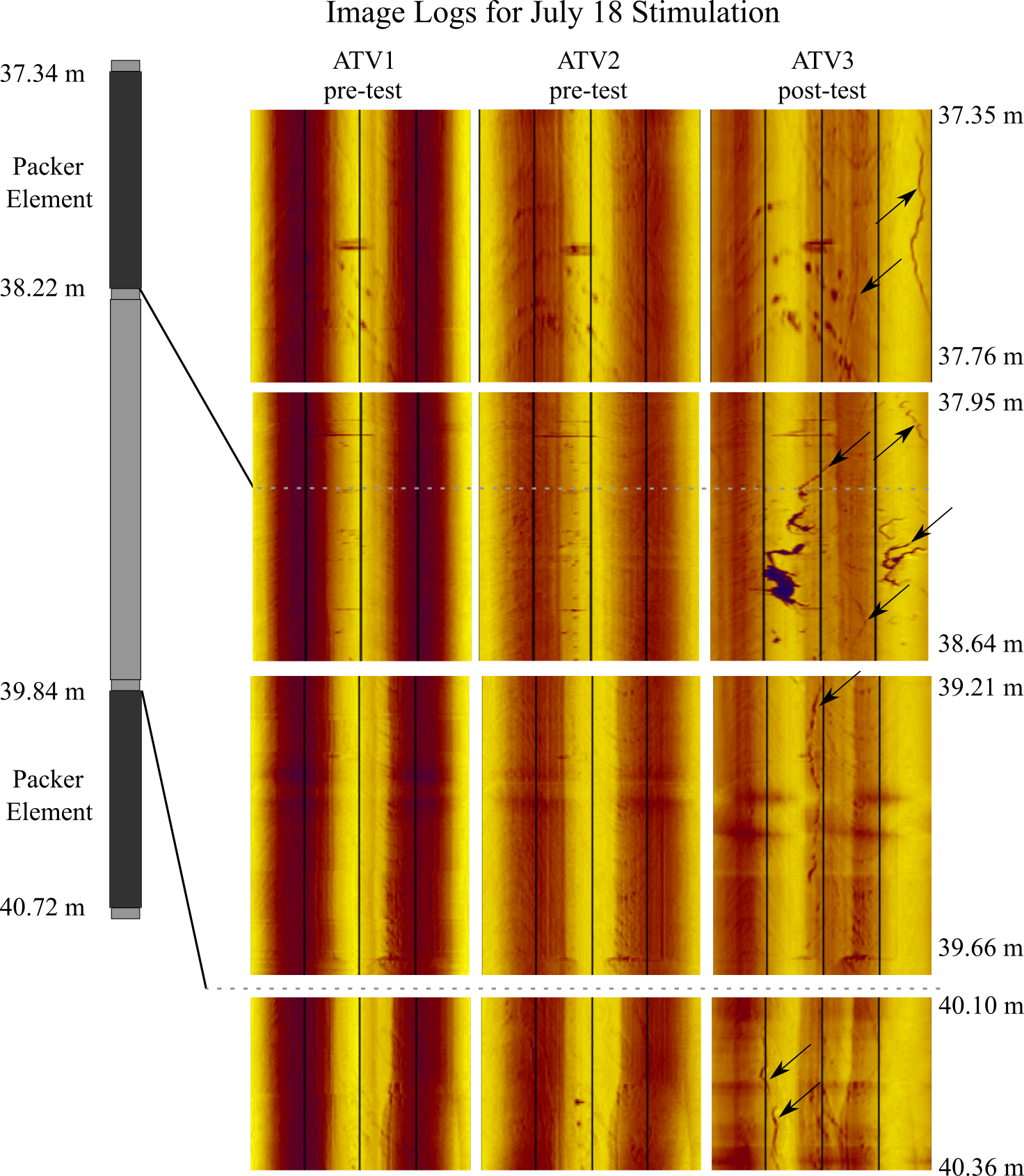}
\caption{\label{fig:fig11} Acoustic (ATV) image logs collected over the zone stimulated on July 18; both ATV1 and ATV2 were collected prior to the stimulation while ATV3 was in January 2020 the stimulation; the left most edge of each log is magnetic North, which is within a few degrees of the top of the borehole}
\end{figure}

Figure \ref{fig:fig11} shows pre- and post-test image logs from the interval tested in the July 18 test. Sub-longitudinal fractures can be observed under the upper packer at 37.4 to 37.5 m, and again from  38 to 38.6 m. These fractures are rotated approximately 10 degrees from the top and bottom of the well with some en echelon fracturing. This suggests that the principal stresses are not well aligned with the well at this location. A longitudinal fracture can be seen with in the pressurized interval, spanning 39.2 to 39.6 m, just below the notch. Longitudinal fractures can also be seen underneath the lower packer from 40.15 to 40.35 m and again at 40.55 to 41.02 m. Similar to the fractures under the upper packer, these appear to be rotated by approximately 10 degrees from the top/bottom of the well.

 After a significant volume of fluid was pumped into this fracture on July 20, 2018. A momentary connection with one of the monitoring wells was made. However, injection was terminated before a connection to the production well was made so as not to further damage the integrity of hte monitoring well. The only information available for the orientation of the induced fracture away from the borehole is from seismic event locations and the thermal anomaly detected in the distributed temperature sensing (DTS) in the monitoring well. As described by \citep{Schoenball2019b}, two planes were fit to this data that have a shallow angle relative to the borehole and therefore seem to deviate significantly from the far-field stress directions inferred elsewhere in the test bed.

\section{Discussion}
\label{sec:discussion}

Fracture initiation pressure at the three stimulation sites varied from a low of 22.1 in the July 18 stimulation, to a high of 25.1 in the May 21 stimulation. The fracture propagation pressure at 200 mL/min varied from a low of 25 MPa in the May 22 stimulation to a high of greater than 31 MPa in the May 22 stimulation.

None of the tests had the type of breakdown or propagation pressure curve that is expected of a radial fracture propagating in a homogeneous stress field; i.e., the propagation pressure of a toughness-dominated radial fracture is expected to decline steadily over time as the fracture radius grows. However, none of the tests showed a significant decline in pressure after fracture initiation. Only the May 22 stimulation had a pressure decline following fracture initiation, and the decline was only 1.2 MPa, and it was followed by a steady increase in pressure to greater than the initial peak pressure.

Pre- and post-stimulation image log analysis for the stimulations at 43.2 m and 50 m leave an ambiguous picture of the nature of the fracture initiation. The results presented above suggest that in the stimulation of the notch at 43.2 m, several closely-spaced oblique fractures formed at the borehole. The possibility that a hydraulic fracture also initiated from the notch at this location cannot be ruled out. Results from the stimulation at the 50 m notch indicate fracture initiation either from the notch, or on natural features that left no signature on the initial image log. Logging after subsequent testing at this location suggests that opening of foliation planes occurred.

For both the stimulations centered at 43.2 m and 50 m, the intersection with both monitoring well E1-OT and the production well, E1-P were along multiple closely-spaced fractures, and the intersections and microseismic events suggest that the fracture plane was within just a few degrees of being perpendicular to the stimulation well.

The closely-spaced fractures induced in the stimulation well during the May 21 stimulation, and the closely-spaced fractures at the intersections seems most consistent with a borehole that is at an oblique angle to the principal stresses, as with the $60^{\circ}$ tests in \citet{Abbas1996}. However, all other data at this site suggest that the borehole is within just a few degrees of the minimum principal stress. 

The complex rock fabric in this test bed is a complicating factor that, given current understanding, is difficult to account for. This fabric contributes both elastic anisotropy, and many planes of weakness in the form of foliations and natural fractures. In some cases the foliation planes are only $20-30^{\circ}$ misaligned from the preferred fracture plane, making them likely to interact with a propagating or initiating hydraulic fracture. It is possible that a fracture could have initiated along foliation planes and left no signature on the image logs \citep{Schwering2020}, however a downhole camera was used to observe flow into the production well, and the inflow does not seem to be along foliation planes. Given the complexity of the folded and strongly foliated rock, it is possible that the average stress is aligned with the direction of the larger scale fracture, but that at the smaller scale the stresses are rotated sufficiently to form en echelon fractures.

\citet{Jeffrey2015} deserves a more in-depth comparison because of its similarity with the present study. In their study a large number of vertical wells were drilled and hydraulically fractured using a straddle packer system similar to the one discussed here. At the test site of the Jeffrey study the minimum principal stress is vertical such that horizontal fracture propagation is preferred, which is transverse to the vertical wells used. As with the present study, radial notches were used in many of the wells to encourage initiation of transverse fractures. Image logs were collected after fracturing and many of the wells showed both longitudinal and transverse fractures. In some wells no fracture could be seen on the image logs, which was taken as an indication that the fracture had initiated from the notch.

In the tests reported in this paper, there was clear evidence of initiation of longitudinal fractures in the pressurized zone of the July 18 test, and underneath the packers for all three tests. In the May 22 test, it appears that after later high pressure testing both longitudinal fractures and fractures along foliation planes oblique to the well formed. Despite the similarities, a primary difference between the present study and the Jeffrey study is that the injection rates were much lower in the present study. Several studies have found that higher viscosity fluids and higher pressurization rates increases the likelihood of initiating both longitudinal and transverse fractures \citep{Weijer1994, Abbas2013}. The observation of fractures in the January 2020 image log near the 50 m notch that were not apparent in the May 2018 image log, support this understanding because higher rate and pressure tests were conducted between the two logs. 

\section{Conclusion}
\label{sec:conclusion}
The results of three hydraulic stimulations at the EGS Collab site at the Sanford Underground Research Facility have been reported. The stimulation borehole was designed to be perpendicular to the minimum principal stress, and the experimental data suggest that this is the case. To help with initiation of a transverse fracture, a radial notch was created at each stimulation location using a cutting tool mounted to a drill string. The notch was designed to be approximately 6 mm deep. An unfortunate consequence of using a notch is that any fracture initiated from the notch is not visible on image logs, leaving ambiguity in the pre- and post-stimulation image log analysis.

Fracture initiation pressure was generally in the range expected with low-rate low-viscosity fluid injection, just slightly higher than the minimum principal stress. The variability in fracture initiation pressure between the three sites seems to be similar to the degree of variability of in situ stress from previous measurements.

Both microseismic data and hydraulic fracture intersections with monitoring and production wells indicate that a large mostly planar fracture was created as a result of these stimulations, and that these planar fractures were within a few degrees of being perpendicular to the stimulation and production wells. Despite the suggestion of a relatively simple, planar fracture, the pressure vs. time trend is not consistent with traditional models of radial hydraulic fracture propagation. Specifically, the propagation pressure had a trend of increasing over time rather than decreasing toward the minimum principal stress as expected by toughness-dominated fracture models.

Another observation that is incompatible with a simple planar fracture model is that when the hydraulic fractures intersected a monitoring borehole and a production borehole, it was via multiple closely-spaced fractures rather than a single fracture. Because the stress direction seems to be very consistent across the test bed, a likely explanation for these results is the complex rock fabric. The relevant rock fabric at this site consists of both closely-space foliations and multiple well-developed natural fracture sets, some of which have considerable native hydraulic conductivity.

\section*{Acknowledgements}
The EGS Collab team is: J. Ajo-Franklin, S.J. Bauer, T. Baumgartner, K. Beckers, D. Blankenship, A. Bonneville, L. Boyd, S.T. Brown, J.A. Burghardt, T. Chen
% ...

This material was based upon work supported by the U.S. Department of Energy, Office of Energy Efficiency and Renewable Energy (EERE), Office of Technology Development, Geothermal Technologies Office, under Award Number DE-AC52-07NA27344 with LLNL, Award Number DE-AC05-76RL01830 with PNNL, and Award Number DE-AC02-05CH11231 with LBNL. Sandia National Laboratories is a multimission laboratory managed and operated by National Technology \& Engineering Solutions of Sandia, LLC, a wholly owned subsidiary of Honeywell International Inc., for the U.S. Department of Energy’s National Nuclear Security Administration under contract DE-NA0003525. This paper describes objective technical results and analysis. Any subjective views or opinions that might be expressed in the paper do not necessarily represent the views of the U.S. Department of Energy or the United States Government. The United States Government retains, and the publisher, by accepting the article for publication, acknowledges that the United States Government retains a non-exclusive, paid-up, irrevocable, world-wide license to publish or reproduce the published form of this manuscript, or allow others to do so, for United States Government purposes. The research supporting this work took place in whole or in part at the Sanford Underground Research Facility in Lead, South Dakota. The assistance of the Sanford Underground Research Facility and its personnel in providing physical access and general logistical and technical support is acknowledged.

\bibliography{CollabGeothermicsReferences}
\bibliographystyle{elsarticle-harv}

\end{document}